\documentclass[12pt]{article}

\usepackage{graphicx}

\begin{document}
\baselineskip=1.2em

\title{Stochasticity and Non-locality of Time}

\author{Toru Ohira
\\ \\
Sony Computer Science Laboratories, Inc., \\
Tokyo, Japan 
141-0022
\\
ohira@csl.sony.co.jp}

\maketitle


\begin{abstract}
We present simple classical dynamical models to 
illustrate the idea of introducing a stochasticity with non-locality into the time variable. 
For stochasticity in time, these models include noise in the time variable but not in the ``space" 
variable, which is opposite to the normal description of stochastic dynamics. 
Similarly with respect to non-locality, we discuss delayed and predictive dynamics which
involve two points separated on the time axis.
With certain combinations of fluctuations and non-locality in time,
we observe a ``resonance'' effect.
This is an effect similar to stochastic resonance, which has been discussed within the 
normal context of stochastic dynamics, but with different mechanisms. 
We discuss how these models may be developed to fit a broader context of generalized dynamical systems
where fluctuations and non-locality are present in both space and time.
\end{abstract}

\clearpage

\section{Introduction}

``Time" is a concept that has drawn a lot of attention from thinkers in 
virtually all disciplines \cite{davies1995}. In particular,  
our ordinary perception is that space and time are not the same, and this difference 
appears in various contemplations of nature. It appears to be the main 
reason for the theory of relativity, which has conceptually brought space and 
time closer to receiving equal treatment, and it continues to fascinate and attract 
thinkers from diverse fields. Moreover, issues such as the ``direction" or 
the ``arrow" of time \cite{savitt1995} and complex time \cite{elnasch1995} 
are current research interests. 

It seems that there are other manifestations of this difference. One is the treatment of noise or fluctuations in 
dynamical systems. Time in dynamical systems, whether they are classical, quantum, or 
relativistic, is commonly viewed as not having stochastic characteristics.
In stochastic dynamical theories, we associate noise 
and fluctuations with only ``space'' variables, such as the position of a 
particle, and not with the time variable. In quantum mechanics, the concept of 
time fluctuation is embodied in the time-energy uncertainty 
principle. However, time is not treated as a dynamical quantum observable, 
and clear understanding of the time-energy uncertainty has yet to be found \cite{Busch2002}. 

Another difference seems to show up in our cognition of non-locality in space and time.
Non-local effects in space are incorporated in physical theories describing wave propagation, 
fields, and so on. 
In quantum mechanics, the issue of spatial non-locality
is more intricate, constituting the backbone of such quantum effects as the Einstein-Podolsky-Rosen
paradox \cite{sakurai85}. With respect to time,
there have been investigations of memory effects in dynamical equations. However,
less attention has been paid to non-locality in time, and behaviors associated with
non-locality in time, such as delay differential equations, are not
yet fully understood \cite{mackey77,cooke82,milton89,milton96}.

Against this background, the main topic of this paper is to consider 
stochasticity and non-locality of time in classical dynamics through a presentation of simple 
models. We discuss delayed and predictive dynamics as examples of non-locality in time.
For stochastitiy, we present a delayed dynamical model with fluctuating time, or stochastic time.
We shall see that this combination of stochasticity and non-locality in time can exhibit
behaviors which are similar to stochastic resonance \cite{wiesenfeld-moss95,bulsara96,gammaitoni98}, which arises through
 a combination of 
oscillating behavior and  ``spatial'' noise and has been studied in variety of fields \cite{mcnamara88,longtin-moss91,collins1995,chapeau2003,lee2003}.

\section{Delayed and Predictive Dynamics}

We start with a consideration of non-locality of time in classical dynamical models.
The general differential equation of the class of dynamics we 
discuss here is as follows. 
\begin{equation}
{dx(t) \over dt} = f(\bar{x}(\bar{t}), x(t)).
\end{equation}

Here, $x$ is a dynamical variable of time $t$, and $f$ is 
the ``dynamical function" governing the dynamics. Its difference from the 
normal dynamical equation appears in $\bar{t}$, which can be either
 in the past or the future, and $t\neq\bar{t}$ in general. In other words, the change in 
$x(t)$ is governed by $f$, not  
its ``current" state $x(t)$, but its state $\bar{x}$ at $\bar{t}$. We can define $\bar{t}$ and
 $\bar{x}$, as well as the 
function $f$, in a variety of ways. In the following, we will present two cases: delayed and predictive dynamics.

Delayed dynamics can be obtained from the general definition by
\begin{equation}
\bar{t} = t-\tau, \quad \bar{x}(\bar{t}) = x(t-\tau).
\end{equation}
Here, $\tau > 0$ is the delay, and the dynamics depend on two
points on the time axis separated by $\tau$. Delayed dynamical equations have been studied
for various applications \cite{mackey77,cooke82,milton89,milton96}.

Predictive dynamics, on the other hand, have recently been proposed \cite{ohira06-1,ohira06-2} and, they take
$\bar{t}$ in the future, i.e., $\bar{t}=t+\eta$. We call $\eta>0$ an ``advance''.
We also need to define the state of
the dynamical variable $x$ at this future point in time. Here, we estimate $x$ such 
that
\begin{equation}
\bar{x}(\bar{t}=t+\eta) =\eta {dx(t) \over dt}+ x(t).
\end{equation}
This prediction is termed ``fixed rate prediction''. Namely, we estimate 
$x$ as the value that would be obtained if the current rate of change 
extends for a duration from the present point to the future point. 
Qualitatively, this is one of the most 
commonly used methods for estimating  population, national debt, and so on. 

We also note that there are studies of equations called ``functional 
differential equations of the advanced type'', or ``advanced functional differential equations''
\cite{kusano1981,kolmanovskii1992,agarwal2004}.
They also are differential equations with advanced arguments, and we
can obtain equations of this class from our general definition by setting,
\begin{equation}
\bar{x}(\bar{t}=t+\eta) =x(t+\eta),
\end{equation}
with suitably chosen boundary conditions.
The predictive dynamical equations differ from this class of equations, as
we allow flexibility in defining $\bar{x}$  based on a prediction scheme.

We shall investigate the properties of these delayed and predictive dynamical models through computer simulations. 
To avoid ambiguity and for simplicity, we will study time-discretized map 
dynamical models, which incorporate the above--mentioned general 
properties of the delayed and predictive dynamical equations.
\begin{equation}
x_{n+1}  = (1-\alpha) x_{n} + f [\bar{x}_{\bar{n}}]
\end{equation}
Here, $\alpha$ is a parameter controlling the rate of change.
For a delayed map with a delay $\tau$, we define
\begin{equation}
\bar{x}_{\bar{n}} \equiv x_{n-\tau},
\end{equation}
while for the predictive map with an advance $\eta$, we have 
\begin{equation}
\bar{x}_{\bar{n}} \equiv  x_{n} + \eta (x_n - x_{n-1}) .
\end{equation}

We choose  the Mackey-Glass function as the dynamical function (Fig. 1), i.e., 
\begin{equation}
f(x) ={ {\beta x} \over {1 + x^{s}}},
\label{mgfunction}
\end{equation}
where $\beta$ and $s$ are parameters.
This function was first proposed for modeling the cell reproduction process and is known to induce chaotic behavior with a large delay \cite{mackey77}. 
\begin{figure}[h]
\begin{center}
\includegraphics[width=.45\textwidth]{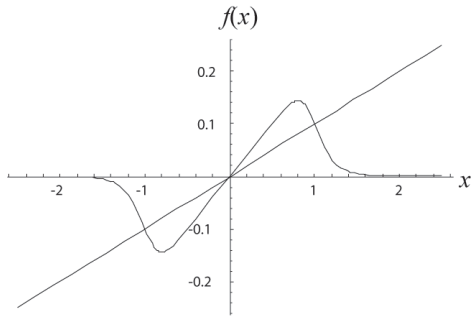}
\caption{
Mackey-Glass function $f(x)$ with $\beta = 0.8$ and $s = 10$. The straight line has 
a slope of $\alpha = 0.1$.
}
\end{center}
\end{figure}

Figure 2 shows examples of computer simulations of the delayed and predictive
cases. The parameters are set so that without a delay or advance, $\tau=\eta=0$, the model 
 monotonically approaches the stable fixed point. The stability of the fixed point is lost 
as $\tau$, or $\eta$, increases, giving rise to complex dynamics.
Thus, non-locality in time can induce a complex behavior in otherwise simple dynamical systems.
\begin{figure}
\begin{center}
\includegraphics[width=.8\textwidth]{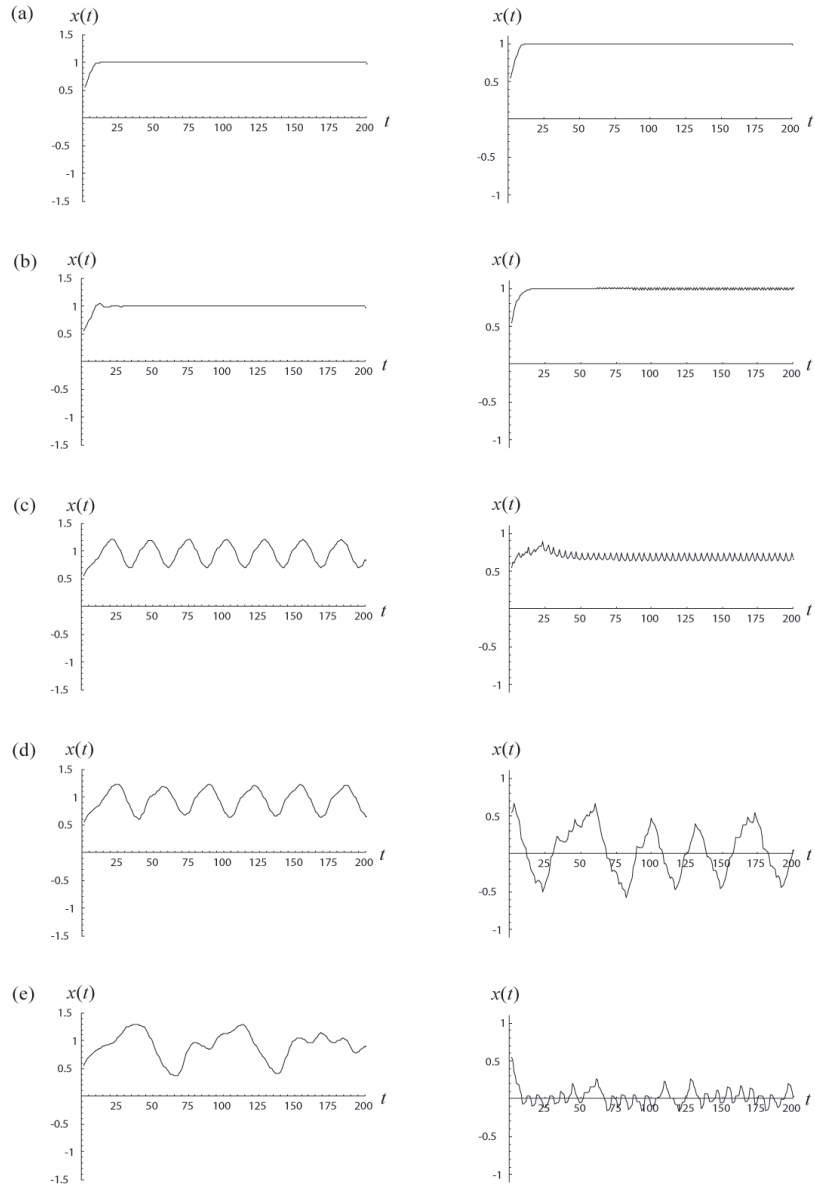}
\caption{
Examples of delayed (left column) and  predictive (right column) dynamics for the
Mackey-Glass map with $\alpha = 0.5$, $\beta = 0.8$, and $n = 10$. 
For delayed dynamics, the initial condition is fixed at $x_0 = 0.5$ for the interval of $(-\tau,0)$.
For predictive dynamics, the initial condition is $x_0 = 0.5$ and $x_1 = (1-\mu) x_{0} + f(x_0)$.
The values of delay and advance $\tau = \eta$ are (a) 0, (b) 2, (c) 8, (d) 10, and (e) 20.
}
\end{center}
\end{figure}

Now, we would like to make a few remarks.
First,
in the case of delayed dynamics, we need to decide on the initial function and delay. Analogously, in predictive dynamics,
the prediction scheme and advance need to be specified. Common to delayed and predictive dynamical systems, both factors  affect the nature
of the dynamics. 

In addition, we can use linear stability analysis on both the delayed and predictive cases. This analysis can
give an estimate of the critical delay or advance at which the stability of the fixed point is lost.
However, the nature of the dynamics beyond these critical points is not yet clearly understood.

For the case of delayed dynamics, with the addition of a suitable ``strength'' of noise, a behavior
similar to stochastic resonance has been obtained \cite{ohira-sato}. This phenomenon, called ``delayed stochastic resonance,'' 
has a different mechanism in the sense that it does not require external oscillatory signals or forces, but
instead it uses a delay as the source of the oscillation to be combined with noise. An analogous resonance phenomenon has been observed
in predictive dynamics with added noise, which is termed ``predictive stochastic resonance''\cite{ohira06-2}.

\section{Stochastic Time}

We now turn our attention to the fluctuation of time, which we term ``stochastic time'' in the context of classical 
dynamical systems. As in non-locality, there are various ways to bring in stochasticity.
We have found that stochastic time combined with delayed dynamics leads to phenomena similar to stochastic resonance.

The general differential equation of the class of delayed dynamics with stochastic time is given as
\begin{equation}
{dx(\bar{t}) \over d\bar{t}} = f(x(\bar{t}), x(\bar{t}-\tau)).
\end{equation}

Here, as in the previous section, $x$ is the dynamical variable of time $t$, and $f$ is 
the ``dynamical function" governing the dynamics. $\tau$ is the delay. The difference from the 
normal delayed dynamical equation appears in $\bar{t}$, which now contains stochastic 
characteristics, and these can be introduced in various ways. We will again focus on the following dynamical 
map system incorporating the basic ideas of the general definition given 
above. 
\begin{eqnarray}
x_{n_{k+1}} & = & f(x_{n_k}, x_{{n_k}-\tau}),\nonumber\\
n_{k+1} & = & n_k + \xi_k
\end{eqnarray}
Here, $\xi_k$ is the stochastic variable which can take either $+1$ or $-1$
with certain probabilities. We associate ``time'' with an integral variable $n$. The dynamics progress by incrementing integer
$k$, and $n$ occasionally ``goes back'' a unit with the occurrence of $\xi=-1$.
Let the probability of $\xi_k =-1$ be $p$ for all $k$, and we set $n_{0}=0$. Then, with $p=0$,
this map naturally reduces to a normal delayed map with $n_k=k$. We update the variable $x_n$ with the larger $k$.
Hence, $x_n$ in the ``past''
could be ``re-written'' as $n$ decreases with probability $p$. 

We can make an analogy of this model with a tele--typewriter or a tape--recorder, which
occasionally moves back on a tape. Figure 3 gives a schematic view. Based on the values of $x_n$ and $x_{n-\tau}$, the recording device writes on
the  tape the values of $x$ at a step, and ``time'' is associated with  positions on the tape.
When there is no fluctuation ($p=0$), the head moves only in one direction on the tape and it records  values of $x$ for a normal delayed dynamics.
With probability $0 < p$, it moves back a unit of ``time'' to
overwrite the value of $x$. The question is how the recorded patterns of $x$ on the tape are affected 
as we change $p$.
\begin{figure}[h]
\begin{center}
\includegraphics[width=.5\textwidth]{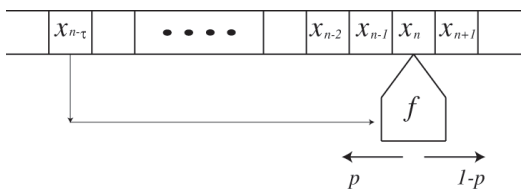}
\caption{Schematic view of the model.}
\end{center}
\end{figure}

We will keep the Mackey-Glass function as the dynamical function, and the
map model becomes
\begin{eqnarray}
x_{n_{k+1}} & = & (1-\alpha) x_{{n_k}} + f(x_{{n_k}-\tau}),\nonumber\\
n_{k+1} & = & n_k + \xi_k,\nonumber\\
f(x) & = & { {\beta x} \over {1 + x^{s}}},
\end{eqnarray}
where $\alpha$, $\beta$, and $s$ are parameters. With both $\alpha < \beta$ positive, and
no stochasticity in time, this map has a stable fixed point
with no delay. Linear stability analysis around the fixed point gives the critical delay $\tau_c$, at which the stability of the fixed point is
lost. 
\begin{equation}
\tau_c \sim {\cos^{-1}({\alpha \over \gamma}) \over \sqrt{\gamma^2 - \alpha^2} }, \quad
(\gamma \equiv \alpha \{ 1-(1-{\alpha \over \beta}) s \})
\end{equation}

A larger delay gives an oscillatory dynamical path.
We have found, through computer simulations, that an interesting behavior
arises when the delay is smaller than this critical delay. The tuned noise in the time
flow gives the system a tendency for oscillatory behavior. In other words,
by adjusting the value of $p$ controlling $\xi$, one induces oscillatory dynamical paths.
Some examples are shown in Figure 4. (The critical delay is $\tau_c \sim 22.5$  with the set
of parameters.) With increasing probability for a time
flow to reverse, i.e., with increasing $p$, we observe oscillatory behavior in the sample 
dynamical path as well as in the corresponding power spectrum.
However, as $p$ increases further, the oscillatory behavior begins to 
deteriorate. In order to see this, we compute the ``signal--to--noise'' ($S/N$) ratio by
using the ratio of the peak height to the background in the spectrum.
 Figure 5 illustrates this change in $S/N$, which reaches a maximum at
an appropriately ``tuned'' value of $p$. 

\begin{figure}
\begin{center}
\includegraphics[width=.98\textwidth]{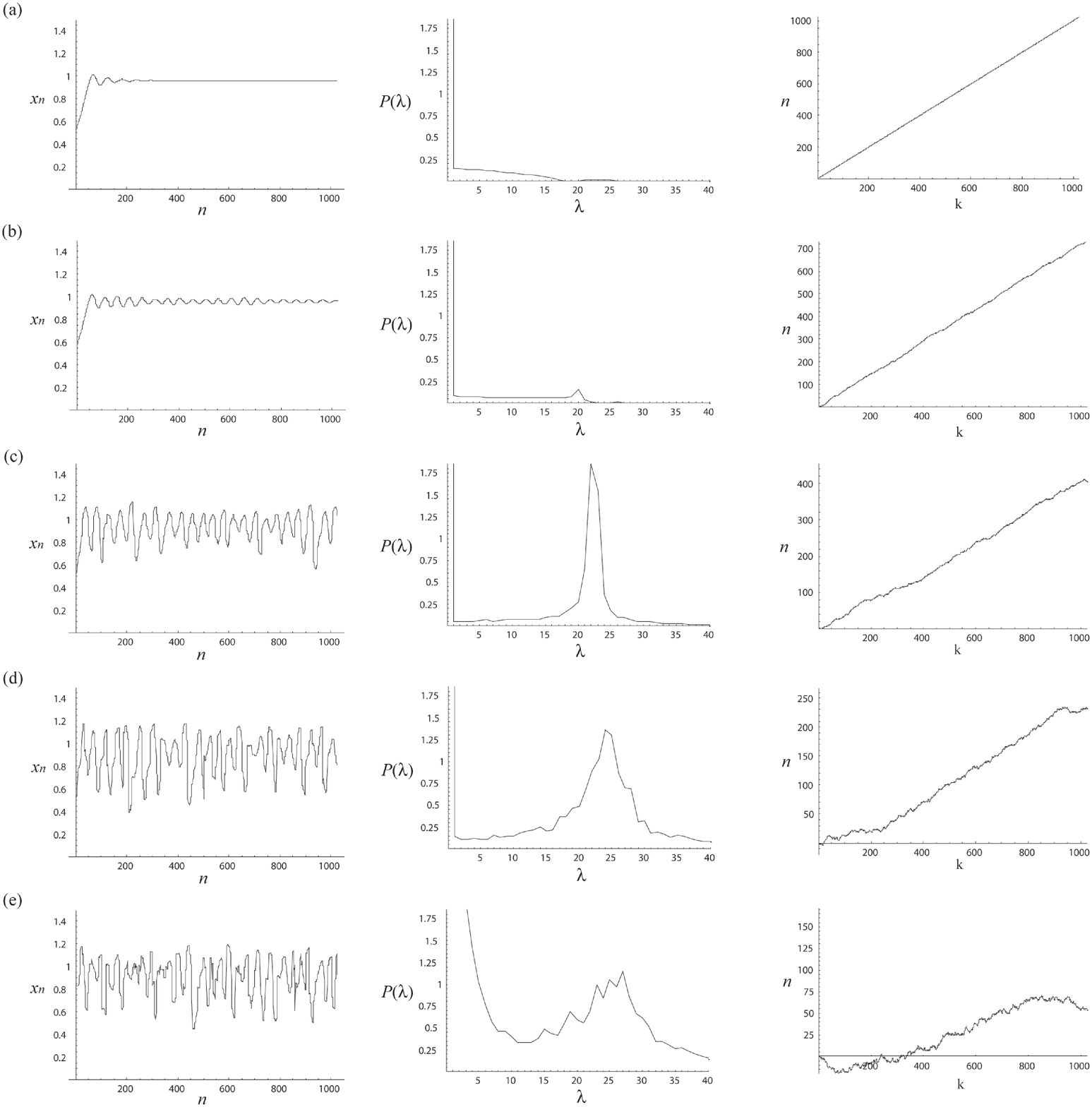}
\caption{Dynamics (left) and power spectrum (middle) of delayed dynamical model with 
stochastic time. (The right column plots the values of stochastic time $n$ as a function of $k$.)
This is an example of dynamics and associated power 
spectrum simulated with the model of Eq. (10) with a variable probability
of stochastic time flow $p$. The 
parameters are  $\alpha = 0.03$, $\beta = 0.05$, $s=10$, and $\tau=15$, and the 
stochastic time flow parameter $p$ is set to (a) $p=0$, (b) $p=0.15$, (c) $p=0.3$, (d) 
$p=0.4$, and (e) $p=0.45$. We used the initial conditions $x_{n} = 0.5 (n\leq 0)$, and
 ${n_0}=0$.
The simulation had $k=10240$ steps, and  the values of $x_n$ for $0 \leq n \leq L$ with
$L=1024$ were recorded at that point. Fifty averages were taken for the power spectrum of this recorded $x_n$. The unit of frequency $\lambda$ is set as ${1 \over L}$, 
and the power $P(\lambda)$ is in arbitrary units. 
}
\end{center}
\end{figure}

\begin{figure}
\begin{center}
\includegraphics[width=.5\textwidth]{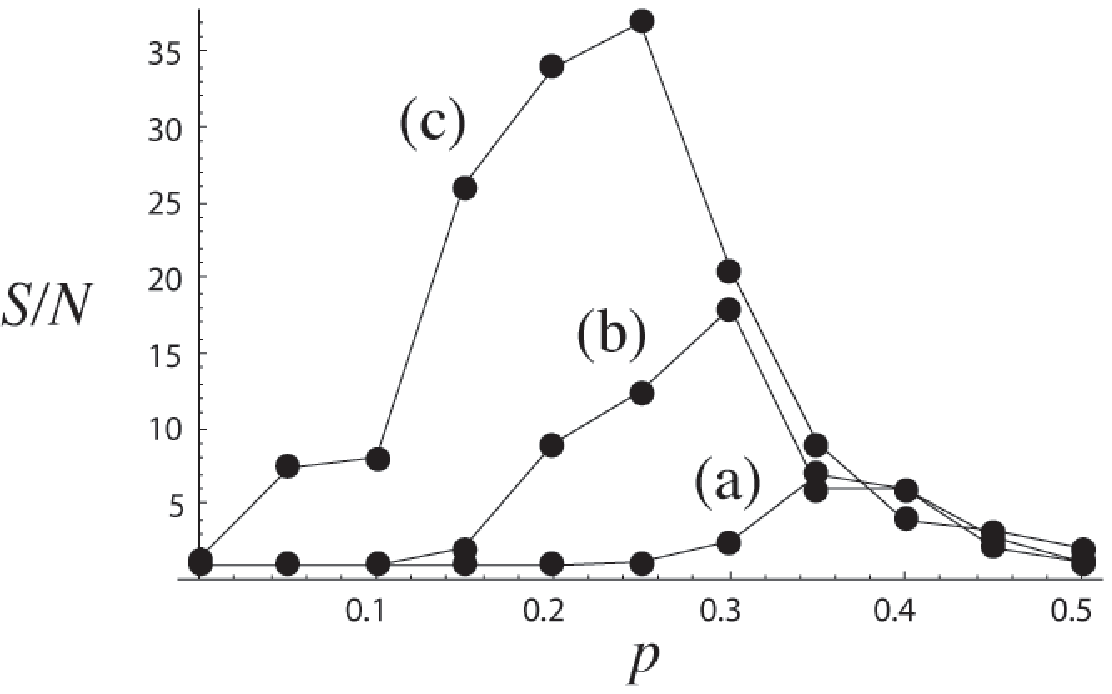}
\caption{Signal--to--noise ratio ${S / N}$ at the peak as a function of the probability  
of stochastic time flow $p$. The parameter settings are the same as in Figure 4 with (a) $\tau=10$, (b) $\tau=15$, and (c) $\tau=20$.
}
\end{center}
\end{figure}

Again, we see a phenomenon which resembles stochastic resonance. A theoretical analysis
of the mechanism of our model is yet to be done.  In particular, we note that the model has an intricate
mixture of
time scales of delay, the oscillation period of $x_n$, and stochastic time.
 These factors are likely to be involved in the time scale analysis \cite{cabrera2002},
but we leave consideration of them for the future. 
On the other hand,
this resonance with stochastic time is clearly of a different type and new. 
We discussed only the Mackey--Glass function for consistency, but the same behavior is
also observed in other delayed dynamical systems, such as ones with a negative feedback function \cite{ohira06-3}.

We have also studied predictive dynamics with stochastic time, but so far, have not found behaviors similar to this resonance with
delayed dynamics. 
Yet, the example above with delayed dynamics indicates
 that a combination of stochasticity and non-locality
in time may lead to entirely new phenomena.

\section{Discussion}

We could extend our model so that we have a picture of dynamical systems with 
stochasticity and non-locality on the time and space axes. The analytical framework and 
tools for such descriptions need to be developed, along with a search for 
appropriate applications. 

An example of an appropriate application of temporal non-locality is  
modeling a stick balancing on a human fingertip. Recent 
experiments have found that most of the observed corrective motions occur on 
shorter time scales than that of the human reaction 
time \cite{cabrera-milton02,cabrera-milton04B,cabrera-etal04}. This may be 
the result of intricate mixtures of physiological delays, predictions, and 
physical fluctuations. Models incorporating special fluctuations and delays 
have been considered, but none have tried to include the effect of prediction.

Another 
direction of development might be to extend the path integral formalism to allow stochastic time paths. 
The question of whether this extension bridges to quantum mechanics and/or leads to 
an alternative understanding of such properties as the time-energy uncertainty 
relations requires further investigation.

Finally, if these models can capture some aspects of reality, particularly with respect to
temporal stochasticity, this resonance 
may be used as an experimental indication for probing fluctuations or 
stochasticity in time. We have previously proposed ``delayed stochastic 
resonance''\cite{ohira-sato}, a resonance that occurs through the interplay of noise and a delay. 
It was theoretically extended \cite{tsimring01}, and recently, it was 
experimentally observed in a solid-sate laser system with a feedback 
loop \cite{masoller}. We leave it for the future to see if an analogous 
experimental test could be developed with respect to stochasticity and non-locality of time.

\end{document}